\documentclass[12pt]{article}
\usepackage[colorlinks=true, urlcolor=navyblue, linkcolor=navyblue, citecolor=navyblue]{hyperref}
\usepackage{graphicx,amsmath,amssymb}
\usepackage{indentfirst}
\usepackage[usenames]{color}
\usepackage{epstopdf}
\usepackage{enumerate}
\usepackage{appendix}
\usepackage{setspace}

\definecolor{navyblue}{rgb}{0,0.08,0.45}
\definecolor{darkred}{rgb}{0.7,0.0,0.0}
\definecolor{darkgreen}{rgb}{0,0.6,0.2}

\begin{document}


\begin{flushright}
{\small SLAC--PUB--16167 \\ \vspace{2pt}}
\end{flushright}

\vspace{20pt}

\begin{center}
 {\bf \large
 The Common Elements of Atomic and Hadronic Physics}
\end{center}

\vspace{20pt}

\centerline{Stanley J. Brodsky}

\vspace{3pt}

\centerline {\it SLAC National Accelerator Laboratory,
Stanford University, Stanford, CA 94309, USA~\footnote{\href{mailto:sjbth@slac.stanford.edu}{\tt sjbth@slac.stanford.edu}}}

\vspace{40pt}

\begin{abstract}

Atomic physics and hadronic physics are both governed by the Yang Mills gauge theory Lagrangian; in fact, Abelian quantum electrodynamics  can be regarded as the zero-color  limit of quantum chromodynamics.  I review a number of areas where the techniques of atomic physics can provide important insight into hadronic eigenstates in QCD.  For example, the Dirac-Coulomb equation, which predicts the spectroscopy and structure of hydrogenic atoms, has an analog in hadron physics in the form of frame-independent  light-front relativistic equations of motion consistent with light-front holography which give a remarkable first approximation to the spectroscopy, dynamics, and structure of light hadrons.  The production of antihydrogen in flight can provide important insight into the dynamics of hadron production in QCD at the amplitude level.  The renormalization scale for the running coupling is unambiguously set in QED;  an analogous procedure  sets the renormalization scales in QCD, leading to scheme-independent scale-fixed predictions.  Conversely, many techniques which have been developed for hadron physics, such as scaling laws, evolution equations, the quark-interchange process and light-front quantization have important applicants for atomic physics and photon science, especially in the relativistic domain.

\end{abstract}

\newpage

\section{Introduction}
\label{intro}
Quantum Electrodynamics -- the fundamental theory of leptons and photons which underlies atomic and molecular physics -- and Quantum Chromodynamics --  the  quark and gluon theory  that underlies hadronic and nuclear physics,   are both governed by the Yang-Mills Lagrangian.  
In fact, in the limit  where the number of colors $N_C$ vanishes,  with $\alpha_s C_F = \alpha$ held fixed ($C_F \equiv (N^2_C- 1) /2 N_C$),  the non-Abelian theory becomes equivalent to Abelian gauge theory~\cite{Brodsky:1997jk}.  This analytic connection between QCD and QED  provides a link between the two fields;   processes  and analyses in QCD at $N_C \to 0$ must be compatible with analogous processes and constraints for QED.  There are, in fact, many examples where particle and atomic physics confront similar theoretical challenges, particularly in the area of relativistic bound-state dynamics.

In this contribution to the EXA2014 proceedings, I will briefly review several topics where the techniques of atomic physics give important insight into the structure of hadrons, the color-singlet bound states of quarks and gluons in QCD.  For example, the Dirac-Coulomb equation, the effective equation that predicts the spectroscopy and structure of hydrogenic atoms has an analog in hadron physics in the form of relativistic frame-independent equations of motion derived from light-front holography~\cite{deTeramond:2008ht}, equations which give a remarkable first approximation to the spectroscopy, dynamics, and structure of light hadrons. Other topics include:
The production of atoms in flight provides a method for computing the formation of hadrons from QCD jets -- ``hadronization at the amplitude level".  The renormalization scale for the running coupling which is unambiguously set in QED leads to a solution for setting  renormalization scales in QCD.  Conversely, many techniques and analyses developed for hadron physics, such as scaling laws, and evolution equations,  have equal utility for atomic physics.  One of the most powerful tools used in hadron physics is light-front quantization~\cite{Brodsky:1997de} -- based on Dirac's ``front form".  Light-Front methods provide many important tools for analyzing the dynamics of atoms in motion and  thus have many  applications of interest for photon science and laser physics.

 \section{Light-Front Quantization}
 
The distributions of electrons within an atom are conventionally determined by  eigenfunctions of the QED Hamiltonian $H \vert  \Psi \rangle = E \vert \Psi \rangle$. 
at fixed time $t.$  However, this traditional method -- called the ``instant form" by Dirac,~\cite{Dirac:1949cp} is  plagued in the relativistic theory by complex vacuum and  casualty-violating effects, as well 
as  by  the fact that the boost of  fixed-$t$ wavefunctions away from the hadron's rest frame is a difficult dynamical problem.  
However, there is an extraordinarily powerful non-perturbative frame-independent alternative -- quantization at fixed light-front (LF) time $\tau = t + z/c = x^+ = x^0 + x^3$ -- the ``front-form" of Dirac.~\cite{Dirac:1949cp}
The constituents of a bound state in a light-front wavefunction are measured (e.g.  in Compton scattering)  at a fixed light-front time $\tau$ -- along the front of a light-wave, as in a flash picture.  In contrast, the constituents of a bound state in an ``instant form: wavefunction must be measured at the same ``instant time" $t$;  this requires the exact synchrony of simultaneous probes.

In the light-front framework, an atom in QED  or a  hadron $H$ in QCD is identified as an eigenstate of the LF  Hamiltonian 
$H_{LF} \vert \Psi_H \rangle = M^2_H \vert \Psi_H \rangle$,   
where  $H_{LF} = P_\mu P^\mu= P^- P^+ -  P^2_\perp$ is the light-front time evolution operator which is derived directly from the Yang-Mills Lagrangian. The eigenvalues of this Heisenberg equation give the complete mass spectrum of the theory. The eigensolution  $|\Psi_H \rangle$  projected on the free Fock basis  provides  the  set of valence and non-valence  light-front Fock state wavefunctions $\Psi_{n/H}(x_i, k_{\perp i}, \lambda_i)$, which describe the bound-state's internal momentum and spin distributions.  
If one quantizes the vector field in light-cone gauge $A^+= A^0 + A^3=0$, the photons and gluons have physical polarization $S^z = \pm 1$;  there are no ghosts, so that  one has a physical interpretation of the bound states in terms of their fundamental constituents and spins.

One can utilize the Mandelstam-Liebbrandt method~\cite{Mandelstam:1982cb,Leibbrandt:1983pj,McCartor:1995dc} to regulate the singularities at $k^+=0$ which appear in LF time-ordered perturbative matrix elements and loop integrals when quantizing in light-cone gauge. 
Alternatively, one can choose to quantize a gauge theory in a covariant gauge such as 
Feynman gauge~\cite{Srivastava:1999gi} and avoid these complications. Other regularization issues for LF quantization are discussed in~\cite{Brodsky:1999xj}.

A remarkable feature of LFWFs is the fact that they are frame
independent; i.e., the form of the LFWF is independent of the
hadron's total momentum $P^+ = P^0 + P^3$ and $P_\perp.$
The boost invariance of  LFWFs contrasts dramatically with the complexity of  boosting the wavefunctions defined at fixed time $t.$~\cite{Brodsky:1968ea}  

Light-front quantization is thus the ideal framework to describe the
structure of hadrons in terms of their quark and gluon degrees of freedom~\cite{Brodsky:1997de}.  The
constituent spin and orbital angular momentum properties of the
hadrons are also encoded in the LFWFs.  
The total  angular momentum projection~\cite{Brodsky:2000ii} 
$J^z = \sum_{i=1}^n  S^z_i + \sum_{i=1}^{n-1} L^z_i$ 
is conserved Fock-state by Fock-state and by every interaction in the LF Hamiltonian~\cite{BURKARDT:2014daa}.
The empirical observation that quarks carry only a small fraction of the nucleon angular momentum highlights the importance of quark orbital angular momentum.  In fact the nucleon's anomalous moment and its Pauli form factor are zero unless the quarks carry nonzero $L^z$~\cite{Brodsky:1980zm}.

Hadron observables,  such as hadronic structure functions, form factors, distribution amplitudes,  GPDs, TMDs, and Wigner distributions can be computed as simple convolutions of light-front wavefunctions (LFWFs) in QCD~\cite{Pasquini:2014fja}.  For example,  one can calculate the electromagnetic and gravitational form factors 
$<p+ q| j^\mu(0)| p>$ and $<p+ q| t^{\mu \nu}(0)| p>$ of a hadron from the Drell-Yan-West formula -- i.e., the overlap of LFWFs.    
The anomalous gravitomagnetic moment $B(0)$ defined from the spin-flip matrix element  
$<p+ q| t^{\mu \nu}(0)| p>$ at $ q\to 0$ vanishes~\cite{Brodsky:2000ii} -- consistent with the equivalence theorem of gravity.
In contrast, in the instant form, the overlap of instant time wavefunctions is not sufficient.  One must also couple the photon probe to currents arising spontaneously from the vacuum which are connected to the hadron's constituents.

The  Wick theorem proves that LF time-ordered perturbation theory is equivalent to the Feynman analysis. However, the LF theory has the advantage that only diagrams with positive $k^+$ momenta appear, LF spin $J^z$ is conserved at every vertex, only three dimensional integrals appear, and unitarity is explicit. A recent application to the Parke-Taylor multi-gluon scattering amplitudes in QCD perturbation theory is given by Cruz-Santiago and Stasto~\cite{Cruz-Santiago:2013vta}.  In addition, the numerator algebra is process independent, allowing for efficient recursive methods.
The LF perturbative computation of the lepton anomalous moment up to order $\alpha^3$ including QED renormalization using the ``alternate denominator method is given in ref. \cite{Brodsky:1973kb}  
The perturbative LFWFs of an electron at order $\alpha$ are given in ref.~\cite{Brodsky:2000ii}
The LFWFs for atomic bound states such as positronium is given in ref. ~\cite{Lepage:1980fj} .

The eigenvalues $M^2_i$ of the LF Hamiltonian $H_{LF} |\Psi_i> = M^2_i |\Psi_i>$, define the LF Heisenberg eigenvalue problem. 
and provide the 
spectroscopy of invariant masses  of a  theory;  the projections of the eigenstates $<n||\psi_i>$ on the free Fock basis $|n>$ define the corresponding frame-independent LF wavefunctions. 
For example, one can solve QCD(1+1) for any choice of colors, quark masses and flavors and obtain the entire set of meson and baryon eigensolutions by matrix diagonalization to high precision using the ``discretized light-cone quantization method (DLCQ)"~\cite{Hornbostel:1988fb}.  Other 1+1 field theories that appear in string theory have been solved by using DLCQ by Klebanov~\cite{Demeterfi:1994bv}. 
A new method~\cite{Vary:2013kma}, called ``basis light-front quantization (BLFQ)", uses the orthonormal basis of eigensolutions of the AdS/QCD effective theory as the basis to diagonalize the full Hamiltonian of QCD(3+1).A comprehensive review is given in ref. ~\cite{Brodsky:1997de}. 

Light-front quantization provides a rigorous method for solving nonperturbative quantum field theories.  The formalism is frame-independent and causal.  
The LF vacuum is trivial up to $k^+=0$ `zero modes'~\cite{Nakawaki:1999ee}  For example, in the LF quantization of the Standard model, the Higgs vacuum expectation value is represented as 
a static scalar background $k^+=0$ field~\cite{Srivastava:2002mw}, analogous to a static Zeeman or Stark Field in QED.  
Recent discussions of the physics of zero modes and the breaking of chiral symmetry in  LF QCD are given in refs.~\cite{Beane:2013oia,Brodsky:2012ku,Brodsky:2010xf}.

Light-Front methods are  directly applicable for describing atomic bound states of QED(3+1)  in both the relativistic and nonrelativistic domains; it is thus particularly useful for atoms in flight since the LFWFs are frame-independent. It also satisfies theorems 
such as cluster decomposition~\cite{Brodsky:1985gs} even for relativistic theory.  The use of  light-front time also provides a natural basis for describing laser interactions  with atoms and molecules since light-front time is measured along the front of a light-wave~\cite{Brodsky:2011fc}  The frame-independence of light-front quantization eliminates the complications from Lorentz boosts as well as vacuum processes~\cite{Brodsky:2008xu,Brodsky:2012ku}. 
Implications for the cosmological constant are discussed in ref.~\cite{Brodsky:2008xu}.

As shown by White et al.~\cite{White:1994tj}, elastic hadron-hadron scattering amplitudes are dominated by quark interchange; e.g., the interchange of the  $u$-quark in $K^+p \to K^+ p$ dominates over gluon exchange.  The amplitude for quark interchange can be written as the overlap of the four incident and final hadronic light-front wavefunctions~\cite{Gunion:1973ex}.  In the non-relativistic limit the LF formula reduces to the standard formula for electron-interchange (spin-exchange) in molecule-molecule scattering.

The analog of intrinsic charm Fock states in hadrons~\cite{Brodsky:1980pb}  such as $|uud c \bar c>$  in the proton is the $|e^+ e^- \mu^+ \mu^->$ Fock state of positronium  which  appears through the cut of the muon-loop light-by-light contribution to the self energy of the positronium eigenstate.   In this Fock state, the muons carry almost all of the momentum of the moving atom since the off-shell virtuality is minimal at equal velocity.  In QED the probability for intrinsic leptons $L \bar L$  exist in positronium scales as $1/m^4_L$, whereas in QCD the probability of intrinsic heavy quarks in the wavefunction of a light hadron scales as  $1/ m^2_Q$ because of its non-Abelian couplings~\cite{Brodsky:1984nx,Franz:2000ee}.

\section{The LF Schr\"odinger Equation}

\begin{figure}
 \begin{center}
\includegraphics[width=12cm]{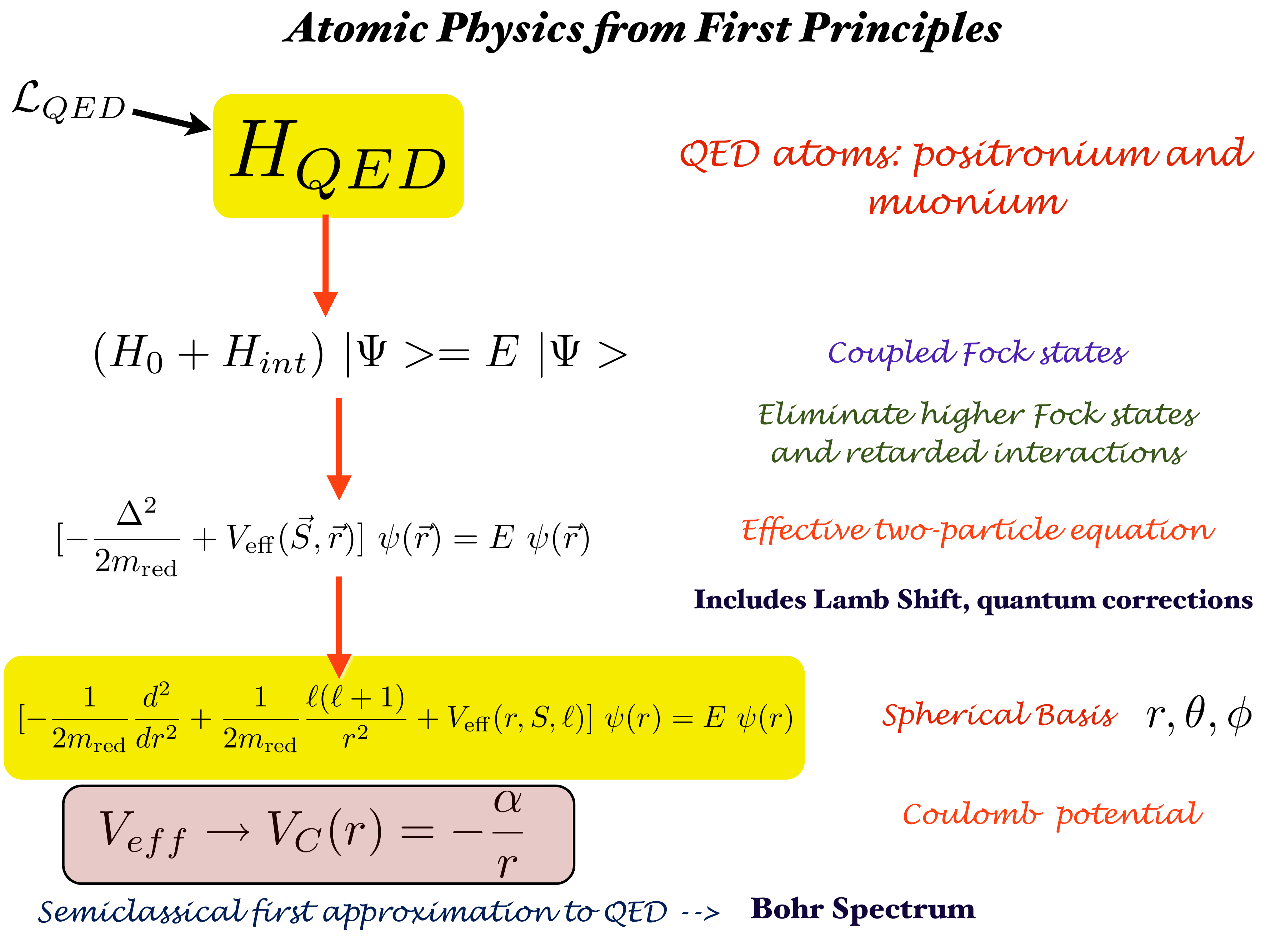} 
\end{center}
\caption{ Reduction of the QED Hamiltonian to the Coulomb Schr\"odinger  Equation }
\label{reductionQED}  
\end{figure}

\begin{figure}
 \begin{center}
\includegraphics[width=12cm]{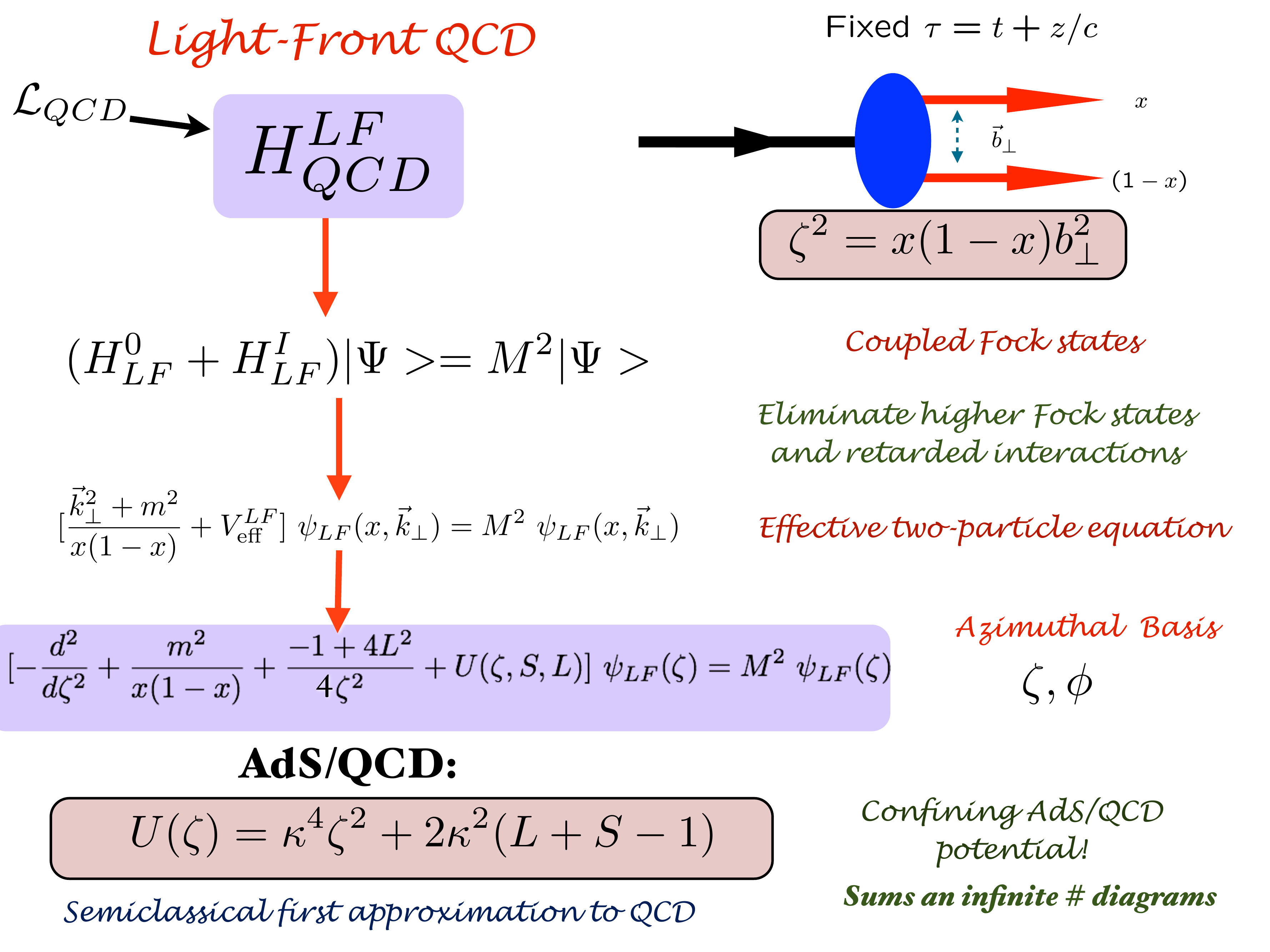}
\end{center}
\caption{ Reduction of the QCD Light-Front Hamiltonian the LF Schr\"odinger  Equation}
\label{reductionQCD}  
\end{figure}

The computation of the precision spectroscopy of hydrogenic atoms in atomic physics is based on reducing the full multi-particle Fock space for QED eigenstates of the QED Hamiltonian to an effective two-body equation. (See Fig. \ref{reductionQED})The effective potential includes the Lamb Shift and other quantum field theoretic effects induced from higher Fock states.
Similarly, it is advantageous to reduce the full multiparticle eigenvalue problem of the LF Hamiltonian for hadronic eigenstates to an effective light-front Schr\"odinger equation  which acts on the valence sector LF wavefunctions of the lowest Fock State of a hadron~\cite{Pauli:1998tf}.  (See Fig. \ref{reductionQCD})
The central problem 
then becomes the derivation of the effective interaction 
$U_{\rm eff}$ which acts only on the valence sector of the theory and has, by definition, the same eigenvalue spectrum as the initial Hamiltonian problem.  In order to carry out this program one must systematically express the higher Fock components as functionals of the lower ones. This  method has the advantage that the Fock space is not truncated, and the symmetries of the Lagrangian are preserved~\cite{Pauli:1998tf}. 
In the exact QCD theory the potential in the Light-Front Schr\"odinger equation (\ref{LFWE}) is determined from the two-particle irreducible $ q \bar q \to q \bar q $ Green's function.  The elimination of the higher Fock states then
leads to an effective interaction $U\left(\zeta^2, J\right)$  for the valence $\vert q \bar q \rangle$ Fock state~\cite{Pauli:1998tf}.

In the case of nonrelativistic QED, one introduces angular coordinates $\theta$ and $\phi$ and the spherical harmonic basis to reduce the 3-dimensional equation to a one dimensional equation in the radial variable $r$. The kinetic energy acquires a term $\ell(\ell+1)/r^2$ for nonzero orbital angular momentum $\ell$. The dominant term in the effective potential is the Coulomb interaction.
(See Fig. \ref{reductionQED})

In the case of the relativistic light-front theory~\cite{Brodsky:2007hb}, the key radial variable is $\zeta^2 = b^2 x(1-x),$ the invariant separation between the quark and  antiquark where $x=k^+/P^+.$  The LF kinetic energy  -- which is  also the invariant mass squared $(p_q+ p_{\bar q})^2$ for a pair of massless quarks -- is 
${k^2_\perp\over x(1-x) } \to -{d^2\over d\zeta_\perp^2 }$.    
One then can introduce the azimuthal angle $\cal \phi$ and a phase factor $\exp{(i L \cal \phi )} $ to obtain a one-dimensional LF Schr\"odinger equation with an extra kinetic energy term $(4 L^2 -1 )/\zeta^2$.  

A novel nonperturbative QCD  approach has been developed which leads to an elegant analytical
and phenomenologically compelling first approximation to the full LF Hamiltonian -- ``Light-Front Holography".~\cite{deTeramond:2008ht}
Light front holographic methods
allow one to project the functional dependence of the wavefunction $\Phi(z)$ computed  in the  fifth dimension of Ant--deSitter space AdS$_5$ to the  hadronic frame-independent light-front wavefunction $\psi(x_i, b_{\perp i})$ in $3+1$ physical space-time. The variable $z $ maps  to the transverse
LF variable $ \zeta(x_i, b_{\perp i})$ which measures the invariant separation of the constituents within a hadron at equal light-front time.
The result is a single-variable light-front Schr\"odinger and Dirac equations which determine the eigenspectrum and the LFWFs of hadrons for general spin and orbital angular momentum.  The transverse coordinate $\zeta$ is closely related to the invariant mass squared  of the constituents in the LFWF  and its off-shellness  in  the LF kinetic energy; it is the natural variable to characterize the hadronic wavefunction.  In fact $\zeta$ is the only variable to appear 
in the relativistic light-front Schr\"odinger equations predicted from 
holographic QCD  in the limit of zero quark masses. 

The effective  potential $U_{\rm eff}$  in the single-variable  LF Schr\"odinger 
equation represents  the complete summation of interactions obtained from the Fock state reduction.  
It is an effective LF equation of motion acting on the lowest valence Fock state 
which encodes the fundamental conformal symmetry of the classical QCD Lagrangian.  

LF Hamiltonian theory provides a rigorous, relativistic, frame-independent framework for solving nonperturbative QCD and understanding the central problem of hadron physics -- color confinement.  The QCD coupling is dimensionless, so the origin of the hadron mass scale for zero quark mass is not  apparent.  However,
if one requires that the effective action which underlies the QCD Lagrangian remains conformally invariant and extends the formalism of de Alfaro, Fubini and Furlan~\cite{de Alfaro:1976je} to light front Hamiltonian theory, the potential $U(\zeta^2)$ has a unique form of a harmonic oscillator potential $\kappa^4 \zeta^2$, and the corresponding dilaton profile of the dual holographic AdS$_5$  model is uniquely determined to be $e^{\kappa^2 z^2} $ and a mass gap arises~\cite{deTeramond:2013it}. 

Thus light-front holography leads to an effective light-front Hamiltonian and relativistic frame-independent wave equation with a unique potential~\cite{deTeramond:2008ht} 
\begin{equation} \label{LFWE}
\left[-\frac{d^2}{d\zeta^2}
- \frac{1 - 4L^2}{4\zeta^2} + U\left(\zeta^2, J\right) \right]
\phi_{n,J,L}(\zeta^2) = 
M^2 \phi_{n,J,L}(\zeta^2).
\end{equation}
The result is a nonperturbative relativistic light-front quantum mechanical wave equation which incorporates color confinement and other essential spectroscopic and dynamical features of hadron physics, including a massless pion for zero quark mass and linear Regge trajectories with the same slope in the radial quantum number $n $and orbital angular momentum $L$.  Only one mass parameter $\kappa$ appears.  Light-front holography thus provides a precise relation between the bound-state amplitudes in the fifth dimension of AdS space and the boost-invariant light-front wavefunctions describing the internal structure of hadrons in physical space-time.
This equation describes the spectrum of mesons as a function of $n$, the number of nodes in $\zeta$, the total angular momentum  $J$, which represent the maximum value of $\vert J^z \vert$, $J = \max \vert J^z \vert$,
A LF Dirac equation for baryons can be derived in a similar way using superconformal quantum mechanics~\cite{deTeramond:2014asa}.   New supersymmetric relations between 
$q \bar q$ mesons with LF angular momentum $L_M$ and baryons with quark-diquark angular momentum $L_B=L_M-1$ appear~\cite{Dosch:2015nwa}.
One finds successful predictions for both hadron spectroscopy and dynamics, including a Regge spectrum with equal slopes in $n$ and $L$ and a zero mass pion at zero quark mass.  A comprehensive review is given in ref.~\cite{Brodsky:2014yha}.

The hadron eigenstates generally have components with different orbital angular momentum; e.g.,  the proton eigenstate in LF holographic QCD  with massless quarks has $L=0$ and $L=1$ light-front Fock components with equal probability.   Higher Fock states with extra quark-anti quark pairs also arise.   The resulting LFWFs then lead to a new range of hadron phenomenology,  including the possibility to compute the hadronization of quark and gluon jets at the amplitude level~\cite{Brodsky:2008tk}.

This approach to hadron physics also predicts the form of the non-perturbative effective coupling for QCD and its $\beta$-function.~\cite{Brodsky:2010ur}.   At low scales the predicted coupling is proportional to 
$\exp{-Q^2/ 4 \kappa^2}$. At high $Q^2$ the coupling falls logarithmically in agreement with asymptotic freedom.  The matching of the two domains then leads to a determination of 
$\Lambda_{\overline {MS}}$ in terms of $\kappa$ which in turn is determined by the $\rho$ or proton mass~\cite{Deur:2014qfa,Brodsky:2014jia}.
The analysis applies to any renormalization scheme.

\section{Production of Exotic Atoms in Flight and Hadronization at the Amplitude Level }
Relativistic antihydrogen was first produced in 1995 at CERN-LEAR~\cite{Baur:1995ck} and at the Fermilab Antiproton Accumulator~\cite{Blanford:1997up}. The incident antiproton beam produces a Bethe-Heitler electron-positron pair in the Coulomb field of a target nucleus $ \bar p Z \to \bar p e^+ e^- Z  \to [\bar p e^+] Z$. The comoving off-shell $\bar p$ and $e^+$ then coalesece into antihydrogen atoms via the Schr\"odinger Coulomb wavefunction which connects the off-shell state to the on-shell anti-atom. The atom is dominantly in its 1S ground state. In principle, one can measure  its ``anti-Lamb-Shift" using  the Robiscoe level-crossing method~\cite{Robiscoe:1965zz}.  

The calculation of the relativistic production of anti-hydrogen is the prototype for computing hadronization at the amplitude level in QCD, now only treated heuristically.
For example,  the  $\Lambda(sud)$ baryon can be produced at high longitudinal momentum fraction $x_F$ in $ p p \to \Lambda X$ reactions by the coalescence of the $ud$ valence quarks of the beam with a comoving strangeness quark. This method can be generalized to produce heavy hadrons such as $\Lambda_c(cud), \Lambda_b,$  double charmed baryons, etc., using the high $ x$ intrinsic heavy quarks which exist in the higher Fock states of  the proton wavefunction~\cite{Brodsky:1981se}.

Consider the production of a $q \bar q$ meson in an $e^+ e^-$ annihilation event.   One first calculates the $T$ matrix element for the production of off-shell quarks and gluons at the amplitude level using light-front time-ordered perturbation theory.  The light-front wavefunction of the meson then converts the off-shell comoving $q \bar q$ pair into the final-state meson.   The confined colored quarks thus never appear on-shell. This first-principle method for forming hadrons in QCD~\cite{Brodsky:2008tk} can replace phenomenological jet hadronization models.
The  light-front wavefunction required for calculating ``hadronization at the amplitude level" ~\cite{Brodsky:2008tk,Brodsky:2009dr} is the frame-independent analog of the Schr\"odinger wavefunction of atomic physics.  It can be determined by solving the Heisenberg matrix $H_{LF}^{QCD}|\Psi_H> = M^2_H |\Psi_H>$ using a method such as discretized light-cone quantization (DLCQ)~\cite{Pauli:1985pv} or using the AdS/QCD approach together with Light-Front Holography~\cite{deTeramond:2008ht}.

It is very interesting to produce  ``true muonium",  the  $[\mu^+ \mu^-]$ bound state.  
Lebed and I~\cite{Brodsky:2009gx} have discussed the QED production and decay mechanisms, such as the electroproduction of relativistic true muonium
below the $\mu^+ \mu^-$  threshold via 
$e^- Z \to [\mu^+ \mu^-]  e^- Z$  or $e^+ e^- \to [\mu^+ \mu^-] \gamma$.  
The APEX  electroproduction experiment~\cite{Abrahamyan:2011gv}, which will search for dark matter candidates at Jefferson Laboratory, could be the first to see this exotic atom. 
Studying the precision spectroscopy of the $[\mu^+ \mu^-]$
atom is important in view of the anomalies seen in the muon $g-2$~\cite{Bennett:2006fi} and the $\mu^- p$ Lamb shift~\cite{Pohl:2010zz}.

``Atomic Alchemy" refers to the transition between a muonic atom into an electronic atom: $(\mu^- Z) \to (e^- Z) {\bar \nu_e }\nu_\mu $ via the weak decay of the bound muon and the subsequent capture of its decay electron.    Greub, Wyler, Munger, and I ~\cite{Greub:1994fp} have shown that such processes provide a laboratory for studying the relativistic high momentum tail of wavefunctions in atomic physics; in addition, they provide a simple toy model for investigating analogous exclusive heavy hadronic decays in quantum chromodynamics such as $B \to \pi e \nu.$

The  QCD analog of a molecule in QCD is a  bound state of heavy quarkonium with a nucleus such as $[J/\psi A]$~\cite{Brodsky:1989jd,Luke:1992tm}. The binding occurs through two-gluon exchange,  the hadronic analog of the Van der Waals interaction. Since the kinetic energy of the $J/\psi$ and the nucleus are both small, one expects to produce these exotic hybrid states at threshold. Examples of nuclear-bound quarkonium are  the $|uud uud s \bar s>$ and $|uud uud c \bar c>$ resonances which apparently appear as intermediate states in $p p \to p p $ elastic exchange. These resonances can account~\cite{Brodsky:1987xw} for the large spin-spin $A_{NN}$ correlations~\cite{Court:1986dh} observed at the strangeness $E_{cm} \simeq 3 $ GeV and $E_{cm} \simeq 5$ GeV and charm thresholds.

\section{Renormalization Scale-Setting in QED and QCD}
A key difficulty in making precise  predictions for perturbative QCD is
the uncertainty in determining the renormalization scale $\mu$ of the
running coupling $\alpha_s(\mu^2)$.  

The fundamental QED coupling    $\alpha(q^2) = \alpha(0) /[1 - \Pi(q^2)]$   is the ``effective charge"~\cite{Grunberg:1982fw} defined from the potential controlling elastic scattering of infinitely heavy leptons~\cite{GellMann:1954fq}.    The function $\Pi(q^2)$  sums the vacuum polarization loops  to all orders.  
The QED $\beta$-function, $\beta(\mu)= d g(\mu)/ d \log \mu$ is analytic~\cite{Brodsky:1998mf} as one successively passes through the vacuum polarization contributions of each intermediate lepton pair. 
The renormalization scale $\mu^2$  in $\alpha(\mu^2)$ is by definition the virtuality of the exchanged photon $\mu^2= q^2.$   For example, in the case of electron-electron elastic scattering, the $t$-channel and $u$-channel one-photon exchange amplitudes are proportional to $\alpha(t) / t$  and $\alpha(u)/u$, respectively. This automatically sums  all vacuum polarization insertions in the renormalized photon propagator.  If one would choose different scales than $t$ and $u$, the evaluation of  an infinite humber of loop contributions would be needed to recover the same result. 
Higher order pQED amplitudes involving multiple-photon exchange each have distinct renormalization scales,  reflecting the  virtualities of the exchanged photons; the renormalization scales can be set in QED at each order so that all terms involving the $\beta$ function (i.e., vacuum polarization contributions) are eliminated.  The resulting series matches that of  the corresponding ``conformal" theory with $\beta=0$.  Renormalon divergences $\alpha^n \beta^n n!$ are eliminated.  Precision tests of QED such as the spectroscopy of muonic atoms depend on the proper choice of renormalization scale -- the
modified muon-nucleus Coulomb potential is precisely
$-Z\alpha(-{\vec q}^{~2})/ {\vec q}^{~2};  $  i.e., $\mu^2=-{\vec
q}^2.$ The renormalization scale in QED is unique.
The Gell-Mann Low coupling~\cite{GellMann:1954fq}  is actually a ``scheme" choice; for example, one could use the $\overline{MS}$ scheme defined from dimensional regularization for pQED.  Different  choices of scheme and effective charges   are related by commensurate scale relations~\cite{Brodsky:1994eh} so that physical observables do not depend on the choice of scheme. To first approximation,
 $\alpha_{GM_L}(Q^2)= \alpha_{\overline{MS}}(e^{5/3} Q^2)[1-   2 \alpha_{\overline{MS}}/\pi +  \cdots] .$   
 
It is conventional to 
``guess: the renormalization scale $\mu$  of the QCD coupling $\alpha_s(\mu^2)$  and its range in pQCD predictions.
The resulting predictions then depend on the choice of the renormalization scheme.
The arbitrary procedure of guessing the renormalization scale and range violates the principle of 
``renormalization group invariance":  physical observables cannot depend on the choice of the renormalization scheme or the initial scale.  Varying the renormalization scale can only expose terms in the pQCD series which are proportional to the $\beta$ function; it is thus an unreliable way to estimate the accuracy of pQCD predictions.

The same principles that are used in QED also unambiguously determine the renormalization scale $\mu_R$ of the running coupling $\alpha_s(\mu_R^2)$ at each order of perturbation theory for QCD.   The essential step: all $\beta$ terms in the pQCD series must be shifted into scales of the running couplings.  At low orders one can identify the $\beta$ terms from the occurrence of $n_F$ terms as proposed in the original BLM paper~\cite{Brodsky:1982gc}; at high orders one can  use the $R_\delta$ method~\cite{Mojaza:2012mf}: one modifies the traditional $\overline{MS}$ dimensional regularization by subtracting an extra constant $\delta$.  The resulting dependence of the pQCD series in powers of $\delta$ unambiguously identifies the $\beta_i$ dependences and the pattern of their occurrence. One then shifts the scales of the QCD coupling at each order of $\alpha^n_s$ to eliminate all $\delta$ and $\beta$ terms. The resulting coefficients of the pQCD series  matches that of the corresponding conformal theory with $\beta=0$. This  ``Principle of Maximum Conformality " (PMC)~\cite{Brodsky:2011ta,Brodsky:2011ig} gives  predictions which rapidly converge and are independent of the choice of renormalization scheme; this is the key principle of renormalization group invariance.  The PMC predictions are also independent of the choice of the initial scale $\mu_R$ to very high accuracy.

There have been many successful applications of the BLM/PMC method. In the case of the forward-backward asymmetry in  $\bar p p \to t \bar t X$ at the Tevatron, the application of the PMC~\cite{Brodsky:2012ik,Wang:2014sua}
eliminates the anomaly reported by CDF and D0 -- the discrepancy  between measurements and pQCD predictions was based on the choice of an erroneous choice of renormalization scale and range. 
As in its QED analog,   $e^+ e^- \to \mu^+ \mu^-$,  the higher  Born  amplitudes which produce  the $\mu^+ \mu^-$ forward backward asymmetry have a smaller renormalization scale than the lowest-order amplitude.  Thus it is essential to assign a different renormalization scale at each order of perturbation theory.  The effective number of flavors $n_f$ is also different at each order.

In summary: The purpose of the running coupling in any gauge theory is to sum all
terms involving the $\beta$ function; in fact, when the
renormalization scale $\mu$ is set properly, all non-conformal
$\beta \ne 0$ terms  in a perturbative expansion arising from
renormalization are summed into the respective running coupling.  The
remaining terms in the perturbative series are then identical to
that of a conformal theory; i.e., the theory with $\beta=0$.  The
divergent ``renormalon" series of order $\alpha_s^n \beta^n n! $
does not appear in the conformal series. Thus,  as in quantum electrodynamics, the
renormalization scale $\mu$ is determined unambiguously by the
``Principle of Maximal Conformality (PMC)" ~\cite{Brodsky:2011ig,Brodsky:2011ta}.  
This is also the principle
underlying BLM scale setting~\cite{Brodsky:1982gc}
An important
feature of the PMC is that its QCD predictions are independent of
the choice of renormalization scheme. The PMC procedure also
agrees with QED scale-setting in the $N_C \to 0$ limit.

The PMC provides a systematic and unambiguous way to set the renormalization scale of any process at each order of PQCD.  An unnecessary  error from theory is eliminated.  The PMC thus allows the LHC to test QCD much more precisely, and the sensitivity of LHC measurements to physics beyond the Standard Model is greatly increased. The PMC is clearly an important advance for LHC physics since it provides an important opportunity to strengthen tests of fundamental theory.

\section{Acknowledgements}

Invited talk, presented by SJB at 
{\it EXA 2014  }  
The International Conference on Exotic Atoms and Related Topics, September 14-19, 2014,
Vienna, Austria.  I thank all of my collaborators whose work has been cited in this report. 
This work  was supported by the Department of Energy contract DE--AC02--76SF00515.    
SLAC-PUB-16167


\newpage

\begin{thebibliography}{}

\bibitem{Brodsky:1997jk} 
  S.~J.~Brodsky and P.~Huet,
  Phys.\ Lett.\ B {\bf 417}, 145 (1998)
  [hep-ph/9707543].


\bibitem{deTeramond:2008ht} 
  G.~F.~de Teramond and S.~J.~Brodsky,
  Phys.\ Rev.\ Lett.\  {\bf 102}, 081601 (2009)
  [arXiv:0809.4899 [hep-ph]].


\bibitem{Brodsky:1997de} 
  S.~J.~Brodsky, H.~C.~Pauli and S.~S.~Pinsky,
  Phys.\ Rept.\  {\bf 301}, 299 (1998)
  [hep-ph/9705477].


\bibitem{Dirac:1949cp} 
  P.~A.~M.~Dirac,
  Rev.\ Mod.\ Phys.\  {\bf 21}, 392 (1949).


\bibitem{Brodsky:1968ea} 
  S.~J.~Brodsky and J.~R.~Primack,
  Annals Phys.\  {\bf 52}, 315 (1969).


\bibitem{Brodsky:2000ii} 
  S.~J.~Brodsky, D.~S.~Hwang, B.~Q.~Ma and I.~Schmidt,
  Nucl.\ Phys.\ B {\bf 593}, 311 (2001)
  [hep-th/0003082].


\bibitem{BURKARDT:2014daa} 
  M.~Burkardt,
  Int.\ J.\ Mod.\ Phys.\ Conf.\ Ser.\  {\bf 25}, 1460029 (2014).


\bibitem{Brodsky:1980zm} 
  S.~J.~Brodsky and S.~D.~Drell,
  Phys.\ Rev.\ D {\bf 22}, 2236 (1980).


\bibitem{Pasquini:2014fja} 
  B.~Pasquini and C.~LorcŽ,
  Few Body Syst.\  {\bf 55}, 287 (2014).


\bibitem{Mandelstam:1982cb} 
  S.~Mandelstam,
  Nucl.\ Phys.\ B {\bf 213}, 149 (1983).


\bibitem{Leibbrandt:1983pj} 
  G.~Leibbrandt,
  Phys.\ Rev.\ D {\bf 29}, 1699 (1984).


\bibitem{McCartor:1995dc} 
  G.~McCartor and D.~G.~Robertson,
  Z.\ Phys.\ C {\bf 68}, 345 (1995)
  [hep-th/9501107].


\bibitem{Srivastava:1999gi} 
  P.~P.~Srivastava and S.~J.~Brodsky,
  Phys.\ Rev.\ D {\bf 61}, 025013 (2000)
  [hep-ph/9906423].


\bibitem{Brodsky:1999xj} 
  S.~J.~Brodsky, J.~R.~Hiller and G.~McCartor,
  Phys.\ Rev.\ D {\bf 60}, 054506 (1999)
  [hep-ph/9903388].


\bibitem{Cruz-Santiago:2013vta} 
  C.~A.~Cruz-Santiago and A.~M.~Stasto,
  Nucl.\ Phys.\ B {\bf 875}, 368 (2013)
  [arXiv:1308.1062 [hep-ph]].


\bibitem{Brodsky:1973kb} 
  S.~J.~Brodsky, R.~Roskies and R.~Suaya,
  Phys.\ Rev.\ D {\bf 8}, 4574 (1973).


\bibitem{Lepage:1980fj} 
  G.~P.~Lepage and S.~J.~Brodsky,
  Phys.\ Rev.\ D {\bf 22}, 2157 (1980).


\bibitem{Hornbostel:1988fb} 
  K.~Hornbostel, S.~J.~Brodsky and H.~C.~Pauli,
  Phys.\ Rev.\ D {\bf 41}, 3814 (1990).


\bibitem{Demeterfi:1994bv} 
  K.~Demeterfi and I.~R.~Klebanov,
  In *Trieste 1993, Proceedings, String theory, gauge theory and quantum gravity '93* 57-95, and Princeton U. - PUPT-1447 (94/01,rec.Feb.) 39 p


\bibitem{Vary:2013kma} 
  J.~P.~Vary, X.~Zhao, A.~Ilderton, H.~Honkanen, P.~Maris and S.~J.~Brodsky,
  Acta Phys.\ Polon.\ Supp.\  {\bf 6}, 257 (2013).


\bibitem{Nakawaki:1999ee} 
  Y.~Nakawaki and G.~McCartor,
  Prog.\ Theor.\ Phys.\  {\bf 103}, 161 (2000)
  [hep-th/9903017].


\bibitem{Srivastava:2002mw} 
  P.~P.~Srivastava and S.~J.~Brodsky,
  Phys.\ Rev.\ D {\bf 66}, 045019 (2002)
  [hep-ph/0202141].


\bibitem{Beane:2013oia} 
  S.~R.~Beane,
  Annals Phys.\  {\bf 337}, 111 (2013)
  [arXiv:1302.1600 [nucl-th]].


\bibitem{Brodsky:2012ku} 
  S.~J.~Brodsky, C.~D.~Roberts, R.~Shrock and P.~C.~Tandy,
  Phys.\ Rev.\ C {\bf 85}, 065202 (2012)
  [arXiv:1202.2376 [nucl-th]].


\bibitem{Brodsky:2010xf} 
  S.~J.~Brodsky, C.~D.~Roberts, R.~Shrock and P.~C.~Tandy,
  Phys.\ Rev.\ C {\bf 82}, 022201 (2010)
  [arXiv:1005.4610 [nucl-th]].


\bibitem{Brodsky:1985gs} 
  S.~J.~Brodsky and C.~R.~Ji,
  Phys.\ Rev.\ D {\bf 33}, 2653 (1986).


\bibitem{Brodsky:2011fc} 
  S.~J.~Brodsky,
  Hyperfine Interact.\  {\bf 209}, 83 (2012)
  [arXiv:1112.0628 [hep-ph]].


\bibitem{Brodsky:2008xu} 
  S.~J.~Brodsky and R.~Shrock,
  Proc.\ Nat.\ Acad.\ Sci.\  {\bf 108}, 45 (2011)
  [arXiv:0803.2554 [hep-th]].


\bibitem{White:1994tj} 
  C.~White, R.~Appel, D.~S.~Barton, G.~Bunce, A.~S.~Carroll, H.~Courant, G.~Fang and S.~Gushue {\it et al.},
  Phys.\ Rev.\ D {\bf 49}, 58 (1994).


\bibitem{Gunion:1973ex} 
  J.~F.~Gunion, S.~J.~Brodsky and R.~Blankenbecler,
  Phys.\ Rev.\ D {\bf 8}, 287 (1973).


\bibitem{Brodsky:1980pb} 
  S.~J.~Brodsky, P.~Hoyer, C.~Peterson and N.~Sakai,
  Phys.\ Lett.\ B {\bf 93}, 451 (1980).


\bibitem{Brodsky:1984nx} 
  S.~J.~Brodsky, J.~C.~Collins, S.~D.~Ellis, J.~F.~Gunion and A.~H.~Mueller,
  DOE/ER/40048-21 P4, SLAC-PUB-15471.


\bibitem{Franz:2000ee} 
  M.~Franz, M.~V.~Polyakov and K.~Goeke,
  Phys.\ Rev.\ D {\bf 62}, 074024 (2000)
  [hep-ph/0002240].


\bibitem{Pauli:1998tf} 
  H.~C.~Pauli,
  Eur.\ Phys.\ J.\ C {\bf 7}, 289 (1999)
  [hep-th/9809005].


\bibitem{Brodsky:2007hb} 
  S.~J.~Brodsky and G.~F.~de Teramond,
  Phys.\ Rev.\ D {\bf 77}, 056007 (2008)
  [arXiv:0707.3859 [hep-ph]].



\bibitem{de Alfaro:1976je} 
  V.~de Alfaro, S.~Fubini and G.~Furlan,
  Nuovo Cim.\ A {\bf 34}, 569 (1976).
  

\bibitem{deTeramond:2013it} 
  G.~F.~de Teramond, H.~G.~Dosch and S.~J.~Brodsky,
  Phys.\ Rev.\ D {\bf 87}, no. 7, 075005 (2013)
  [arXiv:1301.1651 [hep-ph]].


\bibitem{deTeramond:2014asa} 
  G.~F.~de Teramond, H.~G.~Dosch and S.~J.~Brodsky,
  arXiv:1411.5243 [hep-ph].


\bibitem{Dosch:2015nwa} 
  H.~G.~Dosch, G.~F.~de Teramond and S.~J.~Brodsky,
  arXiv:1501.00959 [hep-th].
  
\bibitem{Brodsky:2014yha} 
  S.~J.~Brodsky, G.~F.~de Teramond, H.~G.~Dosch and J.~Erlich,
  arXiv:1407.8131 [hep-ph].




\bibitem{Brodsky:2008tk} 
  S.~J.~Brodsky, G.~F.~de Teramond and R.~Shrock,
  AIP Conf.\ Proc.\  {\bf 1056}, 3 (2008)
  [arXiv:0807.2484 [hep-ph]].


\bibitem{Brodsky:2010ur} 
  S.~J.~Brodsky, G.~F.~de Teramond and A.~Deur,
  Phys.\ Rev.\ D {\bf 81}, 096010 (2010)
  [arXiv:1002.3948 [hep-ph]].


\bibitem{Deur:2014qfa} 
  A.~Deur, S.~J.~Brodsky and G.~F.~de Teramond,
  arXiv:1409.5488 [hep-ph].


\bibitem{Brodsky:2014jia} 
  S.~J.~Brodsky, G.~F.~de TŽramond, A.~Deur and H.~G.~Dosch,
  arXiv:1410.0425 [hep-ph].


\bibitem{Baur:1995ck} 
  G.~Baur, G.~Boero, S.~Brauksiepe, A.~Buzzo, W.~Eyrich, R.~Geyer, D.~Grzonka and J.~Hauffe {\it et al.},
  Phys.\ Lett.\ B {\bf 368}, 251 (1996).


\bibitem{Blanford:1997up} 
  G.~Blanford {\it et al.}  [E862 Collaboration],
  Phys.\ Rev.\ Lett.\  {\bf 80}, 3037 (1998).


\bibitem{Robiscoe:1965zz} 
  R.~T.~Robiscoe,
  Phys.\ Rev.\  {\bf 138}, A22 (1965).


\bibitem{Brodsky:1981se} 
  S.~J.~Brodsky, C.~Peterson and N.~Sakai,
  Phys.\ Rev.\ D {\bf 23}, 2745 (1981).


\bibitem{Brodsky:2009dr} 
  S.~J.~Brodsky and G.~F.~de Teramond,
  arXiv:0901.0770 [hep-ph].


\bibitem{Pauli:1985pv} 
  H.~C.~Pauli and S.~J.~Brodsky,
  Phys.\ Rev.\ D {\bf 32}, 1993 (1985).


\bibitem{Brodsky:2009gx} 
  S.~J.~Brodsky and R.~F.~Lebed,
  Phys.\ Rev.\ Lett.\  {\bf 102}, 213401 (2009)
  [arXiv:0904.2225 [hep-ph]].


\bibitem{Abrahamyan:2011gv} 
  S.~Abrahamyan {\it et al.}  [APEX Collaboration],
  Phys.\ Rev.\ Lett.\  {\bf 107}, 191804 (2011)
  [arXiv:1108.2750 [hep-ex], arXiv:1108.2750 [hep-ex]].


\bibitem{Bennett:2006fi} 
  G.~W.~Bennett {\it et al.}  [Muon G-2 Collaboration],
  Phys.\ Rev.\ D {\bf 73}, 072003 (2006)
  [hep-ex/0602035].


\bibitem{Pohl:2010zz} 
  M.~Pohl and D.~Eichler,
  PoS TEXAS {\bf 2010}, 135 (2010).


\bibitem{Greub:1994fp} 
  C.~Greub, D.~Wyler, S.~J.~Brodsky and C.~T.~Munger,
  Phys.\ Rev.\ D {\bf 52}, 4028 (1995)
  [hep-ph/9405230].


\bibitem{Brodsky:1989jd} 
  S.~J.~Brodsky, I.~A.~Schmidt and G.~F.~de Teramond,
  Phys.\ Rev.\ Lett.\  {\bf 64}, 1011 (1990).


\bibitem{Luke:1992tm} 
  M.~E.~Luke, A.~V.~Manohar and M.~J.~Savage,
  Phys.\ Lett.\ B {\bf 288}, 355 (1992)
  [hep-ph/9204219].


\bibitem{Brodsky:1987xw} 
  S.~J.~Brodsky and G.~F.~de Teramond,
  Phys.\ Rev.\ Lett.\  {\bf 60}, 1924 (1988).


\bibitem{Court:1986dh} 
  G.~R.~Court, D.~G.~Crabb, I.~Gialas, F.~Z.~Khiari, A.~D.~Krisch, A.~M.~T.~Lin, R.~S.~Raymond and R.~R.~Raylman {\it et al.},
  Phys.\ Rev.\ Lett.\  {\bf 57}, 507 (1986).


\bibitem{Grunberg:1982fw} 
  G.~Grunberg,
  Phys.\ Rev.\ D {\bf 29}, 2315 (1984).


\bibitem{GellMann:1954fq} 
  M.~Gell-Mann and F.~E.~Low,
  Phys.\ Rev.\  {\bf 95}, 1300 (1954).


\bibitem{Brodsky:1998mf} 
  S.~J.~Brodsky, M.~S.~Gill, M.~Melles and J.~Rathsman,
  Phys.\ Rev.\ D {\bf 58}, 116006 (1998)
  [hep-ph/9801330].


\bibitem{Brodsky:1994eh} 
  S.~J.~Brodsky and H.~J.~Lu,
  Phys.\ Rev.\ D {\bf 51}, 3652 (1995)
  [hep-ph/9405218].


\bibitem{Brodsky:1982gc} 
  S.~J.~Brodsky, G.~P.~Lepage and P.~B.~Mackenzie,
  Phys.\ Rev.\ D {\bf 28}, 228 (1983).


\bibitem{Mojaza:2012mf} 
  M.~Mojaza, S.~J.~Brodsky and X.~G.~Wu,
  Phys.\ Rev.\ Lett.\  {\bf 110}, 192001 (2013)
  [arXiv:1212.0049 [hep-ph]].


\bibitem{Brodsky:2011ta} 
  S.~J.~Brodsky and X.~G.~Wu,
  Phys.\ Rev.\ D {\bf 85}, 034038 (2012)
  [Erratum-ibid.\ D {\bf 86}, 079903 (2012)]
  [arXiv:1111.6175 [hep-ph]].


\bibitem{Brodsky:2011ig} 
  S.~J.~Brodsky and L.~Di Giustino,
  Phys.\ Rev.\ D {\bf 86}, 085026 (2012)
  [arXiv:1107.0338 [hep-ph]].


\bibitem{Brodsky:2012ik} 
  S.~J.~Brodsky and X.~G.~Wu,
  Phys.\ Rev.\ D {\bf 85}, 114040 (2012)
  [arXiv:1205.1232 [hep-ph]].


\bibitem{Wang:2014sua} 
  S.~Q.~Wang, X.~G.~Wu, Z.~G.~Si and S.~J.~Brodsky,
  Phys.\ Rev.\ D {\bf 90}, no. 11, 114034 (2014)
  [arXiv:1410.1607 [hep-ph]].


 
\end{thebibliography}
\end{document}